\documentstyle[prb,aps,preprint,epsfig]{revtex}
\begin{document}
\draft
\title{Comment on "Strong dependence of the interlayer coupling on the hole mobility
in antiferromagnetic La$_{2-x}$Sr$_x$CuO$_4$ ($x<0.02$)"}
\author{I.~Ya.~Korenblit$^a$, A.~Aharony$^{a,b}$, and O.~Entin-Wohlman$^{a,b}$}
\address{$^a$School of Physics and Astronomy, Raymond and
Beverly Sackler Faculty
of Exact Sciences,\\
Tel Aviv University, Tel Aviv 69978, Israel}
\address{$^b$Department of Physics, Ben Gurion University,
Beer Sheva 84105, Israel}
\date{\today}
\maketitle
\begin{abstract}
Using the experimental data given in the above paper, we show that
-- unlike the effective coupling discussed in this paper --
 the net average antiferromagnetic interlayer coupling in
doped lanthanum cuprates depends only weakly on the doping or on
the temperature. We argue that the effective coupling is
proportional to the square of the staggered magnetization, and
does not supply new information about the origin of the
suppression of the magnetic order in doped samples.
Our analysis
is based on a modified version of the equation describing the
spin-flip transition, which takes into account the decrease of the
staggered moment with temperature and doping.
\\[3ex] xxxxxx
PACS numbers: 74.25.Ha, 74.72.Dn, 75.30.Hx, 75.40.Cx
\end{abstract}

\newpage
The prototype high-temperature superconductor is
La$_{2-x}$Sr$_x$CuO$_4$, which is antiferromagnetic, with the
N\'eel temperature $T_N$ strongly decreasing with increasing
doping for $x \lesssim 0.02$, and superconducting at $x > 0.05$.
Stoichiometric La$_{2}$CuO$_4$ has alternate weak ferromagnetic
moments in the $c$-direction (perpendicular to each CuO$_2$
plane), due to the Dzyaloshinsky-Moriya (DM) interaction, allowed
by its orthorhombic structure. These moments flip into the same
direction (together with a flip of the in-plane staggered
moments), generating a net ferromagnetic moment, at a
temperature-dependent spin-flip (SF) magnetic field $H_c(T)$ along
the $c$ direction.\cite{Thio}

In a recent paper, \cite{H1} H\"{u}cker {\it et al.} studied the
temperature and field dependent spin-flip transition in
La$_{2-x}$Sr$_x$Cu$_{1-z}$Zn$_z$O$_4$. In a separate
paper\cite{H2} they also studied this transition in
La$_{2-x-y}$Eu$_y$Sr$_x$CuO$_4$.
 For each sample, they also
measured the ferromagnetic moment at the SF transition, $M_F(T)$.
The aim of the study was, according to the authors, ``to find out
the primary controlling parameter for the suppression of the 3D AF
order in La$_{2-x}$Sr$_x$CuO$_4$". To this end, they ``have
studied the magnetic interlayer coupling $J_{\perp}$ as a function
of Sr and/or Zn doping" (see p. 1 of their paper). Actually,
however, H\"ucker {\it et al.} calculated the product
\begin{equation}
\Delta E(T)=M_F(T)H_c(T), \label{delE}
\end{equation}
and defined an ``\emph{effective} interplanar coupling", ${\cal
J}_{\perp}(T)= \Delta E(T)/S^2$. It is from the doping and
temperature dependences of this quantity that they tried to find
out  the reason of the suppression of the AF order in doped
lanthanum cuprates.

In this Comment we show that for doped samples the ``effective
interlayer coupling" is not related directly to $J_\perp$ even at
$T=0$. Then we use the experimental data by H\"ucker {\it et al.}
to conclude that in fact $J_\perp$ depends only weakly on $T$ and
$x$. It is probably also independent of $z$. Therefore, $J_\perp$
is {\it not} the primary controlling parameter for the suppression
of the 3D order. The decrease in ${\cal J}_{\perp}(T)$ results
from the decrease in the staggered moment, which is probably
caused by doping dependent changes in the intraplanar correlation
length, due to changes in the intraplanar interactions.

The energy $\Delta E(T)$ should balance the net interplanar
antiferromagnetic exchange coupling energy at the (first order)
flip transition. Therefore, at $T=0$ and $x=z=0$, Thio {\it et
al.} \cite{Thio} equated the energy $\Delta E(0)$ to
$J_{\perp}S^2$, where $S$ represents the ground state staggered
moment per Cu ion, to deduce the
 interlayer
exchange energy
 $J_{\perp}$. Since $J_{\perp}$ results from a
delicate balance between four bonds which couple each Cu ion to
its neighbors in the next plane, one might expect this energy to
depend on doping even at $T=0$.

Our main argument relates to the value of $\Delta E(T)$ at
non-zero temperature and/or doping. As $T$ increases from 0 to
$T_N$, the staggered moment per Cu ion, $M^\dagger$, decreases
from $S$ to zero. A similar decrease may result from the doping,
even at $T=0$. Therefore, the relation $\Delta E(0)=J_\perp S^2$
of Ref. \onlinecite{Thio}, which was also used in Ref.
\onlinecite{H1}, should be generalized into the effective
mean-field interlayer energy given by
\begin{equation}
\Delta E(T)=J_\perp [M^\dagger]^2. \end{equation}
 Thus, ${\cal J}_\perp(T)=J_\perp [M^\dagger/S]^2$,
 and the decrease in ${\cal J}_\perp$ results
 mainly from the decrease in $M^\dagger$. H\"ucker {\it et al.} claimed
that  ${\cal J}_\perp(0)=J_\perp$ (see p. 3 of the paper), and
derived from this relation $J_\perp$ for doped samples. However,
as argued above, this relation holds only for pure samples, when
$M^\dagger=S$. Therefore, the values of ${\cal J}_\perp$ reported
in Table 1 of Ref. \onlinecite{H1} do not represent the interlayer
coupling $J_\perp$.

Note that because of the in-plane and out-of-plane spin
 exchange anisotropy, the staggered moment in lanthanum cuprates is finite
at finite $T$ even at vanishing interlayer coupling, and hence it
need not be very sensitive to the interlayer coupling.
 Its decrease with the increase
of $T$ is due to thermal fluctuations. At relatively high doping,
this decrease may also be due to  stripe formation.\cite{MF}
However, the possibility of stripe formation in the (low doping)
AF ordered region of lanthanum cuprates is still controversial:
The recent experimental results by Gozar {\it et al.}\cite{GD}
exclude the phase separation scenario suggested in Ref.
\onlinecite{MF} for Sr doping in the relevant range $x\leq 0.02$.
At these concentrations, the strong decrease of $M^\dagger$ with
localized hole doping is most probably due to frustration in the
planes.\cite{aa}

Given the mean field DM free energy per site,
 $-4DM^\dagger M_F$, where $D$ is the  DM interaction energy, \cite{TA}
 the ferromagnetic moment is given by
\begin{equation}
  M_F(T)=\chi_\perp 4 D M^\dagger,
  \label{3}
\end{equation}
where $\chi_\perp$ is the transverse ferromagnetic susceptibility.
For the undoped system below $T_N$, $\chi_\perp \approx 1/(8J)$,
where $J$ is the intralayer exchange energy. Therefore, equating
the two expressions for $\Delta E(T)$ yields
\begin{equation}
 \alpha\equiv \frac{H_c(T)}{M_F(T)}=\frac{J_{\perp}}{(4D\chi_\perp)^2}.
  \label{4}
\end{equation}

Using the data from Fig. 3 and Table 1 of Ref. \onlinecite{H1},
we plot in Fig. 1 the ratio $\alpha$ versus $T/T_N$ for different
samples, doped with Sr and  Zn. We also calculated $\alpha$ for Eu
doped samples, using the data from Fig. 24 of Ref.
\onlinecite{H2}.
 Roughly,  $\alpha$ is seen to be essentially the same for
all temperatures and Sr (holes) or single crystal Eu doping (data
for Eu doping are shown only at temperatures where the crystal
remains orthorhombic). The overall slow increase of $\alpha$ with
$T$ can be explained by the decrease of $D$, see Eq. (\ref{4}),
because of the decrease of the orthorhombic distortion with
increasing temperature. \cite{TA} This latter decrease may also
cause a small decrease in $J_\perp$, which is apparently
compensated by the larger decrease in $D$. In any case, these
changes are all small. In contrast, substitution of Cu by Zn
apparently yields somewhat smaller values of $\alpha$. However,
this could still be consistent with no change in $J_\perp$.   It
is known \cite{BH} that in vacancy doped planar isotropic AF
systems the susceptibility $\chi_{\perp}$ diverges at any doping
concentration. Magnetic anisotropy removes the divergency, but the
susceptibility may still be large, since the anisotropy in
lanthanum cuprates is small.
 This increase may also account for the increase of $M_F$
observed in Ref. \onlinecite{H1}, see our Eq. (\ref{3}), and the
decrease of $\alpha$ in Fig. 1. An alternative source for the
decrease of $\alpha$ would involve a doping-dependent octahedral
tilt angle, which would lead to the increase of $D$.\cite{H1} The
same mechanism can explain the decrease of $\alpha$ in
polycrystaline Eu doped samples.


Thus, the values which we deduce for $\alpha$ are consistent with
a scenario in which $J_{\perp}$ essentially does not depend on $T$
or on doping. Since $J_\perp$ represents a net superexchange
energy, which is an average local quantity, this result implies
that fluctuations due to doping average out and have no strong
effect on the measured $J_{\perp}$; the local $J_\perp$ increases
or decreases depending on which sub-lattice is doped. Given that
the average $J_{\perp}$ is constant, the ``effective interlayer
coupling" ${\cal J}_\perp(T)$ does not give more information than
the staggered magnetization $M^\dagger$. The approach of ${\cal
J}_\perp(T)$ to zero when $T$ approaches $T_N$ does not imply that
the interlayer coupling diminishes. Thus, the statement in the
abstract of Ref. \onlinecite{H1}, that the ``interlayer coupling
plays a key role in the suppression of the AF phase", is
unjustified.

The title of Ref. \onlinecite{H1}, which states that the
interlayer coupling (i.e., $J_{\perp}$ rather than ${\cal
J}_\perp$) depends strongly on the hole mobility, is also
misleading. First, as shown above, the change, if any, of
$J_{\perp}$ due to doping is small. Secondly, the paper presents
no direct evidence that the hole mobility has any direct effect on
the magnetic properties of lanthanum cuprates. In contrast, it was
shown that the strong suppression of the AF order by Sr (hole)
doping -- in variance with Zn (vacancy) doping -- can be explained
by the long-range dipole-type magnetic distortion introduced by
localized holes. \cite{GI,CKAE,SK} The theory based on this model
describes quantitatively the phase diagrams of Sr doped as well as
of Sr and Zn doped lanthanum cuprates. \cite{CKAE,KAE} Hence the
attempt of the authors to explain this difference by the effect of
hole mobility is only a suggestion, which has no {\it
quantitative} support. The idea of dynamic magnetic antiphase
boundaries  evoked by the authors to support their statements is,
as noted above, still controversial.

In conclusion, we have shown that  all the available data are
consistent with a constant $J_\perp$, essentially independent of
$T,~x,~y$ and possibly also $z$. Therefore, it is not necessarily
the interlayer coupling which controls the AF order in Sr doped
lanthanum cuprates. In fact, the suppression of the AF order can
be fully explained by the reduction of the in-plane correlation
length with Sr doping, due to frustration.\cite{CKAE,SK}

 We acknowledge useful discussions with B. B\"uchner and with M. H\"ucker,
 and support from
the US-Israel Binational Science Foundation (BSF).

\begin{figure}
\includegraphics[width=8cm]{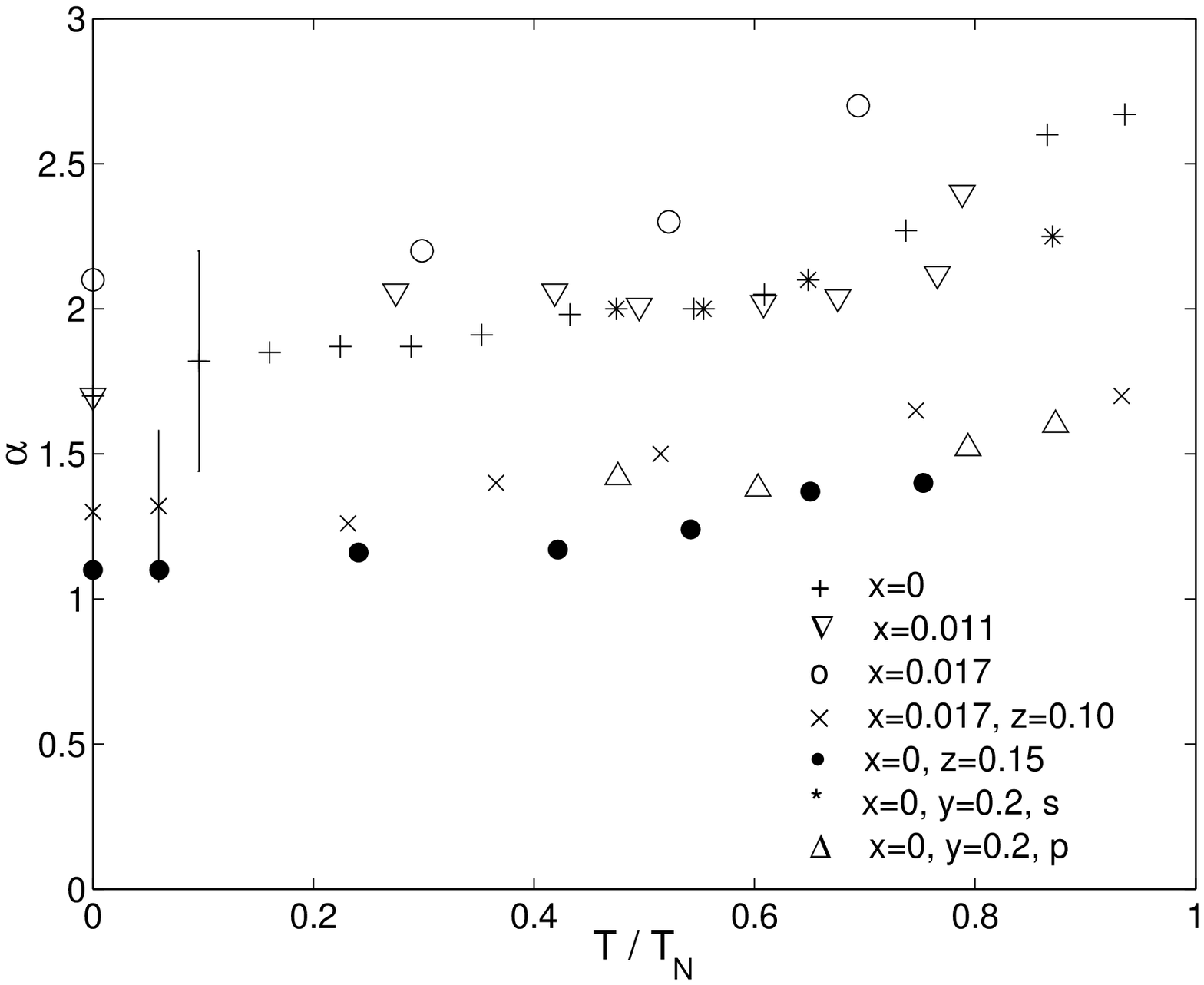}
\caption{The ratio $\alpha=H_c/M_F$ versus temperature  for
lanthanum cuprate doped with Sr (concentration $x$), Zn ($z$) and
Eu ($y$). The experimental data are taken from Refs. [2] and [3].
The typical error bars are from Table 1 in Ref. [2]. ``s" and ``p"
stand for single and polycrystal samples. }
\end{figure}
\end{document}